\documentclass[manuscript, screen, acm]{acmart}
\AtBeginDocument{%
  \providecommand\BibTeX{{%
    \normalfont B\kern-0.5em{\scshape i\kern-0.25em b}\kern-0.8em\TeX}}}

\setcopyright{acmcopyright}
\copyrightyear{2023}
\acmYear{2023}
\acmDOI{}
\acmConference[CSCW User AI Auditing 2023]{CSCW User AI Auditing 2023}{Oct 15, 2023}{MN, USA}

\begin{document}

\title{A Democratic Platform for Engaging with Disabled Community in Generative AI Development}

\author{Deepak Giri}
\email{dgiri@iu.edu}
\affiliation{%
  \institution{Indiana University Indianapolis}
  \country{United States}
}

\author{Erin Brady}
\email{brady@iupui.edu}
\affiliation{%
  \institution{Indiana University Indianapolis}
  \country{United States}
}

\renewcommand{\shortauthors}{Giri and Brady}


\begin{abstract}
  Artificial Intelligence (AI) systems, especially generative AI technologies are becoming more relevant in our society. Tools like ChatGPT are being used by members of the disabled community e.g., Autistic people may use it to help compose emails. The growing impact and popularity of generative AI tools have prompted us to examine their relevance within the disabled community. The design and development phases often neglect this marginalized group, leading to inaccurate predictions and unfair discrimination directed towards them. This could result from bias in data sets, algorithms, and systems at various phases of creation and implementation. This workshop paper proposes a platform to involve the disabled community while building generative AI systems. With this platform, our aim is to gain insight into the factors that contribute to bias in the outputs generated by generative AI when used by the disabled community. Furthermore, we expect to comprehend which algorithmic factors are the main contributors to the output's incorrectness or irrelevancy. The proposed platform calls on both disabled and non-disabled people from various geographical and cultural backgrounds to collaborate asynchronously and remotely in a democratic approach to decision-making.
\end{abstract}

\ccsdesc[500]{Human-centered computing~Accessibility, Collaborative and social computing systems and tools, Visualization systems and tools}

\keywords{Accessibility, Inclusive Design, AI Bias, Social Computing}

\maketitle

\section{Introduction and Related Work}
Generative AI tools like ChatGPT have been used as conversational companions by people on the autistic spectrum to assist them in learning social dynamics and advance their communication abilities with neurotypical people \cite{noauthor_for_nodate}. Although there are some positive uses of generative AI tools by members of the disabled community \cite{leos_5_2023}, there is mounting evidence that these models may produce content that is harmful to them and their particular viewpoints \cite{venkit_study_2022}. AI systems are not neutral when making decisions or generating content. Research has identified ways in which these systems may discriminate against people of a particular origin or skin tone as a consequence of erroneous assumptions \cite{bianchi_easily_2023,deshpande_toxicity_2023}. Much AI bias research and reporting has focused on race and gender, but there has been less attention paid to AI bias and disability. To fill this research void, we are proposing a platform to assist researchers in carrying out investigations of disability bias within the field of generative AI.

As it is critical to develop knowledge of what constitutes disability bias in generative AI  systems and what does not, we think that these questions are appropriate for input from a larger public. Attitudes towards disabilities in a society are influenced by culture. While some cultures are highly inclusive of disabilities and believe it is a natural phenomenon or a social construction, other cultures might view it more negatively, instigating stigma and prejudice against the disabled. Even things like the appropriate terminology to use when talking about different types of disabilities varies greatly across geographies and cultures and it is important for a generative AI system to be aware of that in order to produce relevant and unbiased outputs. For example, South Asian countries consider the term "differently-abled” to address the disabled population whereas in the West the term “disabled” is used, and different communities have different opinions of person-first or identity-first language describing disability. Levels of accessibility, infrastructure and support services vary significantly across the world. Generative AI systems need to be mindful of that and produce content that is relevant to users. With all these factors mentioned, it is necessary to seek broader public understanding to build consensus on the questions of interest. 

With the biases reported in content generated by existing generative AI systems, it is evident that AI systems do not always represent the experiences of marginalized individuals \cite{noauthor_problems_nodate}. It is necessary to democratize the way AI companies operate, because important choices are heavily influenced by a few board members who do not adequately represent the general public \cite{noauthor_towards_nodate}. Oftentimes when generative AI labs are in a race to deploy their systems, they are unaware of the risks and biases associated with their models. The problem are magnified when the systems are deployed on a large scale and their output starts impacting the disabled community directly. The first challenge for labs and developers is to identify the biases that originate from generative AI systems. With our platform, we aim to understand what constitutes a disability bias in generative AI, by gaining consensus on the acceptability of the model’s output. The insights would lay the groundwork for bias identification strategies that development teams could utilize in the premature evaluation of their systems for disability bias. 

There is also a need to revise the conventional data sampling methods in AI research institutions. Disabled groups are typically underrepresented in training data due to their limited involvement in the product development process. We believe that our platform would open up the scope of participation among the disabled in generative AI systems building. With the insights generated through our platform, we aim to increase the advocacy of members of the disabled community by ensuring they are involved  early in the product life cycle. This could transform the existing workings of AI labs and lead toward a more equitable future for all.

In this position paper, we describe a proposed platform to understand disability bias in AI through publics led by  disabled participants. We hope that by participating in this workshop, we will be able to get feedback on the design and identify collaborators for further research.

\section{Methodology}
We have the following research questions in relation to the platform:

\begin{itemize}
        \item \textbf{RQ1:} How does consensus differ between expert-mediated publics and independent publics?
        \item \textbf{RQ2:} Does expert intervention affect participant outlook in expert-mediated publics over time?
    \end{itemize}

We have structured our approach into 3 parts. Part 1 would consist of activities such as platform selection/creation, recruiting experts, prompt development, instrument development and participant recruitment. Part 2 would involve “publics” creation, educating participants about generative AI, running consensus-building activities and gathering algorithmic awareness. The final part would involve an analysis of the data captured during part 2 and mapping the information based on various explored parameters.

\subsection{Part 1 activities:}

\subsubsection{Platform selection/creation:}

The platform should guarantee the security of the data collected throughout the consensus-building part, as well as the privacy of the participants. A major functionality required to support the platform is polling. It should allow admins/experts to define poll questions and response options. Further, enabling admins/experts to specify the duration of the poll and any restrictions. Accessibility features such as keyboard usability, compatibility with screen readers, visual contrast, colour accessibility, and the ability to customize font size and text are crucial.

\subsubsection{Expert recruitment:}

Experts, either with  lived experience of disability or working in the fields of accessibility and social justice, need to be recruited. Academic and professional networks could be utilized to gather interested experts. Researchers should make sure that the experts are recruited from diverse geographical locations so that contextual and cultural relevance in the process is maintained.

\subsubsection{Prompt and instrument development:}

The experts would serve as key personnel in defining relevant themes of discussions and creating a number of prompts surrounding relevant questions of interest for which researchers need to seek deliberate discussions among the participants. Additionally, experts would be responsible for co-creating instruments that would collect participants' initial outlooks on disability bias in generative AI tools, evaluate their knowledge about generative AI tools and gather their algorithmic awareness. Following the development of relevant prompts and instruments by experts, the materials will be chosen based on higher interrater reliability ratings.

\subsubsection{Participant recruitment:}

The platform calls for both disabled and non-disabled individuals as it comprises a representative sample of the population of the affected and influenced public \cite{rowe_public_2000}. Determining who participates is one of the key factors in democratic decision-making, and volunteer-based participation is widely acceptable \cite{lee_webuildai_2019}. Disability among participants could range from visual, hearing, mobility and cognition. A secondary persona of participants, such as family, medical professionals, researchers, special educators, etc., who have close contact with disabled individuals but are not themselves disabled, is something researchers could include while using this platform. Furthermore, the platform calls for participation from various geographic locations to maintain cultural relevance. Disability advocacy organizations and online communities would be sought out to gain participation in the study. Information will be collected from participants, including age, gender, geographical location, familiarity with generative AI tools, outlook on disability bias in AI, required accommodations, disability status quo and type.

\subsection{Part 2 activities:}

\subsubsection{Publics creation:}

Once relevant participants are selected, they should be clustered into smaller “publics” based on their geographic location. The platform proposes clustering based on geographic locations in order to gain participant response inside every publics in a synchronous fashion unaffected by time zones. Diversity in discussions would be achieved by equally clustering participants with varied outlooks towards disability bias in AI. Platform requires half of publics to be expert governed and the remaining half to operate independently.

\subsubsection{Educating about generative AI and evaluating learning:}

Since public participation can be challenging when scientific and technical issues are being debated that call for certain levels of expertise \cite{gilman_beyond_nodate}. The platform requires participants to be educated about the functioning of generative AI tools. They would be introduced to a set of knowledge materials about the topic and would be evaluated through a post-knowledge quiz. 

\subsubsection{Consensus-building activities:}

In a fixed span of time, each publics would be presented with a prompt and the output generated from the generative AI tool. The participants would be asked to vote among three options i.e., the output is biased, unbaised or they are not sure about the it. Also, they would be asked to support their vote with a 100-word justification. Another interesting approach could be using justification statements as choices and creating a collective response system enabling a form of ‘generative voting’ - where both the ‘votes’ and the choices of what to vote on are provided by the participants \cite{ovadya_generative_2023}. Publics which are governed by an expert would receive an opinion on the output from the expert (once all the participants have put in their individual responses). Expert-governed publics would be asked to recast their votes after considering the expert opinions. Participants would be asked for an updated justification statement with a reason why they did or didn’t change their response based on the expert outlook. Once the second round of voting is done in the expert-governed publics, the participants would be shown the distribution of the 2nd round of voting and the updated justification statements. They would then be asked to select a justification statement that best represents their opinion. Vote distribution and highest voted justification statement from one publics would be shared with another randomly selected publics. Prior studies have uncovered that visual support helps individuals identify points of disagreement and leads to effective final evaluation \cite{liu_consensus_2018}. Both types of publics would be asked for an updated justification statement with a reason why they did or didn’t change their response based on the vote distribution and highest voted justification statement from corresponding publics. Distributions of the last round of votes among all publics would be shared with each individual publics.

\subsubsection{Gathering algorithmic awareness:}

Part 2 would be concluded by accessing participants' algorithmic awareness through a set of questions that would identify their understanding on the functioning of the tool and their satisfaction with the kinds of output that algorithms could make based on their experience in consensus-building activities.

\section{Open Questions}

 We believe participating in the workshop will allow us to refine our study design through feedback from the other participants.  Below are some open questions we are still resolving, where we might gain insights from other attendees of the workshop.

\subsection{Analysis}

Through the collected data, we intend to understand how the expert intervention affected participant outlooks in expert-mediated publics over time,  how consensus differs from expert-mediated publics when compared to independent publics, and themes of agreement and disagreement. 

\subsection{Achieving Consensus on Contested Subjects}

Even though we are considering numerous voting and deliberation rounds using expert and intergroup viewpoints, the majority still decides on the acceptability of the generative AI’s outcome, and the decision's correctness may be disputed in some circumstances. However, when considering this from a democratic perspective, it is important to recognize that this limitation can arise as a natural consequence. The platform requires to have an adequate representation of both disabled and non-disabled people, but given regional and cultural differences, it might be challenging to get to the right consensus in some circumstances.

\subsection{Expert Bias}

The influence of experts’ opinions in the voting process in the expert-mediated publics may create biases in decision-making. We already have some ways in mind to lessen this limitation. The first tactic is to compile an expert pool that is diversified in terms of geography, culture, academic and professional backgrounds, etc. Incorporating experts with diverse perspectives can aid in reducing the danger of homogeneous biases and encourage a more impartial and thorough decision-making process. Another strategy involves ensuring that experts present their viewpoints in a balanced manner, discussing both pros and cons, while maintaining impartiality. The final intervention to tackle expert bias should be implemented on the participant’s end. Every time a participant recasts their vote during the second round of voting in the expert-mediated publics, they are required to justify their new or unchanged position. These justifications allow us to determine whether or not expert opinions actually biased the voting.

\subsection{Ensuring Participant Demographics are Accurate}

There are instances when able-bodied participants pretend to be disabled to enjoy certain benefits \cite{dorfman_fear_2019}. Since our platform advocates for asynchronous remote collaboration, researchers would not be able to identify malicious players who would frequently claim these benefits by misrepresenting themselves as disabled. We are still searching for a strategy which will allow us to find participants without risking their privacy by making them validate their disability status.

\bibliographystyle{acm}
\bibliography{sample-base}

\appendix
\end{document}